\documentclass[aps]{revtex4}
\usepackage{amsmath}
\usepackage{epsfig}
\usepackage{float}
\usepackage{dcolumn} 

\begin{document}
\title{\bf One dimensional scattering from  two-piece rising potentials: a new avenue of resonances}  
\author{Zafar Ahmed$^{1}$, Shashin Pavaskar$^{2}$, Lakshmi Prakash$^{3}$}
\affiliation{$^1$Nuclear Physics Division, Bhabha Atomic Research Centre, Mumbai, India \\ $^2$National Institute of Technology, Surathkal, Mangalore, 575025, India \\ $^3$University of Texas, Austin, TX, 78705, USA}
\email{1: zahmed@barc.gov.in, 2: spshashin3@gmail.com, 3:lprakash@utexas.edu} 
\date{\today}
\begin{abstract} 
We study scattering from potentials that rise monotonically on one side; this is generally avoided. We report that resonant states  are absent in such potentials when they are smooth and single-piece
having less than three real turning points (like in the cases of Morse oscillator, exponential and linear potentials). But when these potentials are made two-piece, resonances can occur. We further show that rising potentials next to a  well/step/barrier are rich models of multiple  resonances (Gamow's decaying states) in one- dimension. We use linear, parabolic and exponential profiles as rising part and find complex-energy poles, ${\cal E}_n=E_n-i\Gamma_n/2$ $(\Gamma_n > 0)$, in the reflection amplitude (s-matrix). The appearance of peaks in Wigner's (reflection) time-delay at $E=\epsilon_n$ (close to $E_n$) and spatial catastrophe in the eigenfunction confirm the existence of resonances and meta-stable states in these systems.
\end{abstract}
\maketitle

In one dimension, a potential having (at least) three real turning points at an energy (say, $0< E < V_b)$ can entail metastable states which are characterized by discrete complex energy eigenvalues  $({\cal E}_n=E_n-i\Gamma_n/2, ~\Gamma_n > 0)$. These potentials consist of a well next to a barrier of finite height ($V_b$) and width. The potential $V_{\lambda}(x)=x^2/4-\lambda x^3$ [1] is the  well known example. A rigid wall near a finite barrier is another class of potentials wherein the position of rigid wall itself acts as one real turning point. Ginocchio's versatile exactly solvable potential [2] in a parametric regime becomes a four real turning point system  (a well  surrounded by two finite side barriers) to bear both  metastable ($E_n<V_b$) and resonance $(E_n> V_b)$ states.

In one-dimension, consider a particle in a zero potential domain approaching a rigid wall at $x=a$ from left. Its wave function will be given by  $\psi(x)=A e^{ikx} + B e^{-ikx}$. As it has to vanish at $x=a$, we have $r(E)=B/A = -e^{2ika}$. Here $k=\sqrt{{2mE}}/\hbar$ and $r(E)$ is called reflection amplitude. Its modulus is 1 signifying total reflection from the rigid wall.
The role of reflection amplitude $r(E)$ in one-dimension is like that of $S(E)$ in three-dimensions. 
Now let us ask as to what happens if the rigid wall is replaced by a rising repulsive potential, say $V(x)=V_0 e^{2x/a}$. Schr{\"o}dinger reflection and transmission through this exponential is well studied for $V_0<0$ [3]. For $V_0>0$, one would intuitively claim full reflection at this potential and leave it at that without further investigation. 

Recently, inspired by complex energy eigenvalues in the cubic potential $V(x)=-x^3$ [4], the issue of scattering from a rising  potential has been initiated [5] with the study of an odd parabolic potential. It turns out that for the  rising potential (say for $x>0$), one can actually demand $\psi(\infty) \sim 0$.  For $x\sim -\infty$, on the other hand, one can seek a linear combination of reflected and transmitted wave solutions of the Schr{\"o}dinger equation as per the potential for $x<0$. Unlike the usual reciprocal one-dimensional scattering, the scattering here is essentially one-sided, from left to right if $V(x)$ is rising for $x>0$ and vice-versa. One can then find the reflection amplitude $r(E)$ justifying the intuitive result that $|r(E)|^2=1$. Subsequently, one can extract the complex energy  ($E_n-i\Gamma_n/2, ~\Gamma_n > 0)$ poles of $r(E)= e^{i\theta(E)}$ (zeros of $A(E)$); they cause maxima in the Wigner's time delay $\tau(E)=\hbar \frac {d \theta(E)} {dE}$ [6]. These discrete complex energy states are called Gamow's decaying (time-wise) states. But spatially, they oscillate and grow asymptotically on either one or both sides depending on the potential profile. This behavior of Gamow's states is called catastrophe which should by now [6] be accepted and highlighted as the essence of Gamow's states (also known as Siegert states [7]). This provides a definite mathematical procedure for obtaining  the much sought after resonances and metastable states (mostly, the former is used for both). Solving Schr{\"o}dinger equation by complex scaling of co-ordinate [8] is yet another method of finding resonances.

\begin{figure}[h]
\centering
\includegraphics {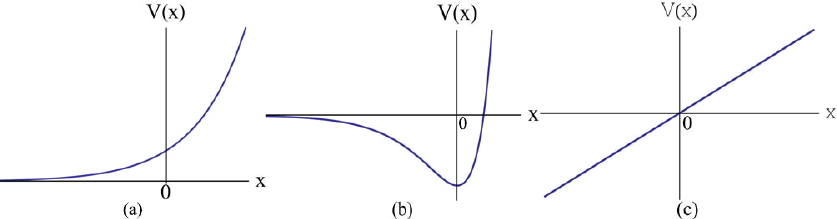}
\caption{Schematic depiction of smooth one-piece potentials having
at most two real turning points --- a: exponential{\bf(S0.1)}, b: Morse{\bf(S0.2)}, c: linear{\bf(S0.3)}. These potentials, we show, do not entail resonances.}
\end{figure}

Here, we consider scattering from rising potentials. We show that resonances do not occur in smooth, one-piece potentials such as those in Fig 1(a,b,c) (See {\bf S0.1}, {\bf S0.2}, {\bf S0.3} below). However, they are  found when the potential is made two-piece such as shown in Figs. 2-4  (See {\bf S1}, {\bf S2}, {\bf S3} below); these at times may be broad resonances. Regardless, they discretize energy in the positive continuum. In addition, we illustrate good, thin resonances for a rising potential (linear, parabolic, and exponential profiles) next to a  well/step/barrier  (See Figs. 5,6,8, see {\bf S4} through {\bf S7} below). In this new avenue, we attribute the occurrence of resonances to the rising potential and its two-piece nature (irrespective of the number of real turning points). 

Earlier, in interesting analyses [9] of rectangular and Dirac delta potentials in  a semi-harmonic background,  discrete complex spectra have been found. However, this  has been unduly attributed  to the semi-harmonic (half-parabolic) potential in particular. It is much in the same way as the complex eigenvalues  of $V_{\lambda}(x)=x^2/4-\lambda x^3$ [1] were attributed to the parabolic (harmonic) part, but now we know [4]  that $-bx^3$  itself supports non-real discrete spectrum. Here we would like to remark that the 
the complex eigenvalues ($E_n <V_b$) in $V_\lambda(x)$ [1] are the metastable states due to three turning point nature of the potentials and those with $E_n >V_b$ owe it to the rising nature of the potential on one side $(x \sim -\infty)$.

Historically, a particle subject to a one-dimensional potential  governed by Schr{\"o}dinger equation
\begin{equation}
\frac{d^2 \psi(x)}{dx^2}+[k^2-\frac{2m}{\hbar^2} V(x)] \psi(x)=0, \quad  k=\sqrt{\frac{2mE}{\hbar^2}}.
\end{equation}
has been at the heart of many phenomena  of the micro-world and continues to be so even today. In this Letter, we solve (1) analytically for various one dimensional systems {\bf S0-S7} which essentially have a rising part in them.
For each system, we calculate the reflection amplitude ($r(E)$), its complex energy poles (${\cal E}_n$), and  Wigner's (reflection) time-delay ($\tau(E)$). Let us define $p,q,s$, for use in sequel 
\begin{small} 
\begin{equation}
p=\sqrt{\frac{2m[E+V']}{\hbar^2}}, ~ q=\sqrt{\frac{2m[E-V_0]}{\hbar^2}}, ~ s=\sqrt{\frac{2mV_0 c^2}{\hbar^2}}.
\end{equation}
\end{small}\\
\noindent
{\bf S0: One-piece rising potentials}\\
Here, we consider an exponential potential ($V_E(x)$) and the well-known Morse oscillator potential ($V_M(x)$).\\ \\
{\bf S0.1:} ${\bf V_E(x)=V_0 e^{2x/c}:}$ {\bf (see Fig. 1a)}\\
The Eq. (1) for $V_E(x)$ can be transformed to Modified cylindrical Bessel equation [10]. We seek the modified Bessel function of second kind $\psi(x)=K_{ikc}(se^{x/c})$ as the physically acceptable  solution. This vanishes for $x \sim \infty$ since [10] $K_{\nu}(z) \sim \sqrt{\frac{\pi}{2z}} e^{-z}\rightarrow 0$. Further, the identity $K_\nu(z)=\frac{I_{-\nu}(z)-I_{\nu}(z)}{\sin \nu \pi}$ [10],  and $I_{\nu}(z)\approx \frac{(z/2)^{\nu}}{\Gamma(1+\nu)}, z\sim 0$ [10], help us to write
\begin{eqnarray}
\psi(x) \sim \left\{ \begin{array}{lcr}
\sqrt{\pi/2s}~ e^{-[x/(2c)+se^{x/c}]}, \quad x\sim \infty \\
-(i\pi kc)^{-1}([(s/2)^{ikc} \Gamma(1-ikc)] e^{ikx}  \\ +[(s/2)^{-ikc} \Gamma(1+ikc)] e^{-ikx}). \quad x\sim -\infty \\
\end{array}
\right.
\end{eqnarray}
thereby giving reflection amplitude as
\begin{equation}
r(E)= -(s/2)^{-2ikc} \left (\frac {\Gamma (1+ikc)}{\Gamma(1-ikc)} \right ).
\end{equation}
It can be readily checked that the all the poles of $r(E)$ are
$ikc=-(n+1)$. These are unphysical as they give rise to a false discrete spectrum $(E_n=-(1+n)^2 \frac{\hbar^2}{2mc^2})$. Hence, no complex energy resonances or peaks in $\tau(E)$ can be found (not shown here). We would like to remark that the wave function (see the dotted line in Fig. 7(c) below) converges to zero on the right and is a combination of incident and reflected waves on the left (as in Eq. (3)). This will be the typical behavior of scattering at a real positive energy (above the well, if any) for all systems (${\bf S0-S7})$ of potentials discussed here (see Figs. 1-6,8). \\ \\
{\bf S0.2: Morse Oscillator potential (Fig. 1(b))} \\${\bf V_M(x)=V_0(e^{2x/c}-2e^{x/c})}$\\
Morse oscillator is a well known exactly solvable potential
having simple explicit expression for discrete 
energy eigenvalues [11,12]. Also the inverted form of this is called Morse barrier which possesses simple and exact expressions [13] for reflection and transmission coefficients. Studying of scattering (positive energy states) from $V_M(x)$ has been avoided as it rises monotonically at positive large distances. Here, we proceed to find the reflection amplitude, $r(E)$, for Morse oscillator.

Inserting $V_M(x)$ in (1), introducing $z=2se^{x/c}$, and using the well known transformation  [11,12]: $\psi(z)= z^{ikc} e^{-z/2} W(z)$;
we notice that $W(z)$ satisfies the Confluent hyper-geometric 
equation (CHE): $zW''(z)+(C-z)W'(z)-A W(z)=0$ [9]. This in our case reads as
\begin{equation}
zW''(z)+(2ikc+1-z)W'(z)-(-s+ikc+\frac{1}{2})W(z)=0.
\end{equation}
The CHE, a second order linear differential equation, has got
two pairs of linearly independent solutions: $(w_1,w_2)$ and $(w_3,w_4)$ [10]. Each one of these pairs is expressible as a linear combination of members from the other pair. $w_1$ is usually chosen as the appropriate solution for bound states. For scattering states $w_3=U(A,C;z)$ ($U$ also denoted as $\psi$ which we avoid here) is appropriate. We discard its other linearly independent partner $w_4$ as it would diverge for asymptotically large positive values.  These considerations lead us to write 
\begin{equation}
\psi(x)=z^{ikc} e^{-z/2} U(-s+ikc+1/2, 1+2ikc; 2se^{x/c}).
\end{equation}
Noting that $U(A,C;z)\sim z^{-A}$ for $z\sim \infty$, we find
$\psi(x) \sim ~(2s)^{s-1/2} \exp[-\{se^{x/c}-sx/c+x/(2c)\}] \rightarrow 0$ $\mbox{as}~~ x \rightarrow \infty$.
Next, we consider the identity[9]
\begin{equation}
w_3=\frac{\Gamma(1-C)}{\Gamma(1+A-C)} w_1+ \frac{\Gamma(C-1)}{\Gamma(A)} w_2,
\end{equation}
where $w_1=~_1F_1(A,C; z), w_2=z^{1-C} ~_1F_1(1+A-C, 2-C; z)$ are Confluent hyper-geometric series (CHS). Since CHS tends to 1 as $z \rightarrow 0$ (equivalently $x\rightarrow -\infty $), the asymptotic behavior of $\psi(x)$
is nothing but a combination of the incident and reflected waves. Using (6,7) we can now write 
\begin{small}
\begin{equation}
\psi(x)\sim (2s)^{ikc} \frac {\Gamma(-2ikc)e^{ikx}}{\Gamma(1/2-s-ikc)} + (2s)^{-ikc} \frac {\Gamma(2ikc)e^{-ikx}}{\Gamma(1/2-s+ikc)}  
\end{equation}
\end{small}
for $x\rightarrow -\infty$. Finally, we extract 
\begin{equation}
r(E)=(2s)^{-2ikc}\frac{\Gamma(2ikc)}{\Gamma(-2ikc)} \frac{\Gamma(1/2-s-ikc)}{\Gamma(1/2-s+ikc)}.
\end{equation}
It readily follows that $|r(E)|^2=1$. The physical poles of
$r(E)$ are obtained by noting that $\Gamma(-n)=\infty$ for $n = 0,1,2,3$ and so on. We get $1/2-s-ikc=-n$ (such that $n + 1/2 < s$), which gives only the well known spectrum of Morse oscillator (Fig. 1(b)) as negative discrete eigenvalues: $E_n=(n+1/2-s)^2 \frac{\hbar^2}{2mc^2}$ [10,11].
The negative energy poles due to $\Gamma(2ip)$ are known to be unphysical. Thus, the Morse oscillator is devoid of resonances; it yields  structureless $\tau(E)$.  
\\ \\
{\bf S1: Two-piece exponential potential} \\
Now we  make the exponential potential $V_E(x)$  of {\bf S0.1} two-piece as $V(x)= V_0 e^{x/c}, ~ x \le 0;~  V_0 e^{x/d}, ~ x>0$. $V(x)$ is continuous but non-differentiable at $x=0$. Earlier, two-piece wells and semi-infinite (step) potentials have been found [14,15] to have a single deep minimum in reflectivity as a function of energy. This is opposed to the usual result of monotonically decreasing reflectivity. Again, we expect some striking difference in $\tau(E)$ due to the two-piece nature.
\begin{figure}[h]
\centering
\includegraphics{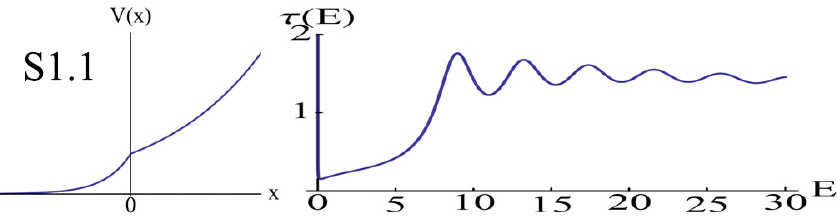}
\includegraphics{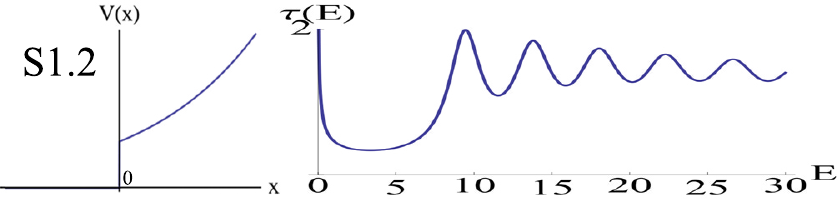}
\caption{Schematic depiction of potential, $V(x)$, on the left and the corresponding time-delay, $\tau(E)$, on the right for system {\bf S1}. For ${\cal E}_n$, $\epsilon_n$, and other details, see Table 1.}
\end{figure}
\noindent
As discussed above in {\bf S0.1}, the left hand solution of (1)
can be expressed in terms of modified cylindrical Bessel functions.
For $x<0$, we seek $\psi(x) = A (s/2)^{-ikc} \Gamma(1+ikc) I_{ikc}(s e^{x/c}) + B (s/2)^{ikc} \Gamma(1-ikc) I_{-ikc}(s e^{x/c})$, and for $x>0$ we take $\psi(x) = C K_{ikd}(s\zeta e^{x/d})$. Using
these solutions and introducing $ \zeta = d/c$, we obtain 
\begin{eqnarray}
r(E)=-(s/2)^{-2ikc}   \frac {\Gamma (1+ikc)}{\Gamma(1-ikc)}
 \left [ \frac{I_{ikc}(s) K'_{ikd}(s\zeta)-I'_{ikc}(s) K_{ikd}(s\zeta)}{I_{-ikc}(s) K'_{ikd}(s\zeta)-I'_{-ikc}(s) K_{ikd}(s\zeta)} \right].
\end{eqnarray}
When $c=d$, both the numerator and denominator in the square bracket are Wronskian functions: $[I_{\nu}(z), K_{\nu}(z)]=1/z$ [10] (real here). These cancel out, thereby giving us (4) back. But when $c\neq d$, $K_{i\nu}(z)$ is real for
real $\nu$ and $z$; the square bracket is uni-modular. Hence, it changes only the phase of $r(E)$ as compared to the phase of $r(E)$ in (4). $r(E)$ in (10) gives  rise to complex energy poles and corresponding peaks in time-delay, $\tau(E)$ (see Fig. 2 and Table 1). This can be seen as a direct consequence of making the potential two-piece.\\ \\
{\bf S2: Two-piece linear potential}\\
\begin{figure}[H]
\centering
\includegraphics{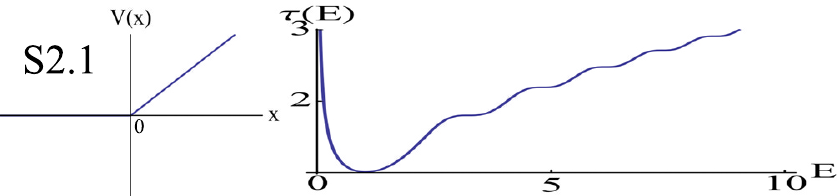}
\includegraphics{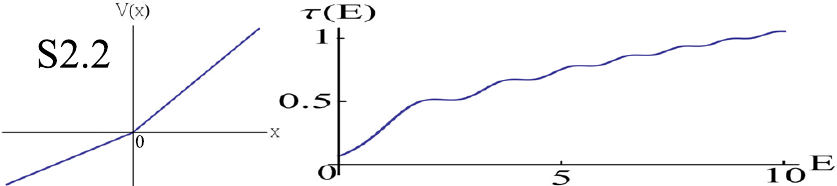}
\caption{The same as in FIG. 2 for system {\bf S2}; See Table 1.}
\end{figure} 
\noindent
Consider $V(x)=hx$, $x<0$; $V(x)=g x$, $x\ge 0$. This potential in Eq. (1) is exactly solvable [10,12] in terms of two linearly independent solutions $\mbox{Ai}(u(x)), \mbox{Bi}(u(x))$ called Airy functions [9]. For the right, we seek $\psi(x)=C \mbox{Ai}(u(x))$, where $u(x)= {2m \over \hbar^2}\left(\frac{gx - E}{g^{2/3}}\right).$ When $x \rightarrow \infty$, $\psi(x)\sim C'u^{-1/4} e^{-2 u^{3/2}/3} \rightarrow 0$. On the left, we use two interesting 
linear combinations: $F_{\mp}(x)=\sqrt{2\pi}[\mbox{Bi}(v(x))\mp i\mbox{Ai}(v(x))]$ [10], where $v(x)= {2m \over \hbar^2}\left(\frac{hx - E}{h^{2/3}}\right).$ Next we notice that $F_{\mp}(x) \sim \sqrt{\pi}\phi(v)e^{\mp i \chi(v)}$, where $\chi(v) \sim 2(-v)^{3/2}/3$. So   when $x \rightarrow -\infty$,  $F_{\mp}$ behave as incident and reflected waves respectively. Thus, we seek $\psi(x)=A F_{-}(v(x)) + B F_{+}(v(x))$ for $x<0$. We get
\begin{equation}
r(E)=-\frac{\eta \mbox{Ai}'(\bar u)[\mbox{Bi}(\bar v)-i\mbox{Ai}(\bar v)]-\mbox{Ai}(\bar u)[\mbox{Bi}'(\bar v)-i\mbox{Ai}'(\bar v)}{\eta \mbox{Ai}'(\bar u)[\mbox{Bi}(\bar v)+i\mbox{Ai}(\bar v)]-\mbox{Ai}(\bar u)[\mbox{Bi}'(\bar v)+i\mbox{Ai}'(\bar v)},
\end{equation}
where $\bar u= -E/g^{2/3}, \bar v= -E/ h^{2/3}$ and $\eta=(g/h)^{1/3}$.
The primes denote differentiations with respect to the parameter in parenthesis. It can be readily checked that when $g=h$, numerator and denominator in (11) become Wronskian $[\mbox{Ai},\mbox{Bi}]=1/\pi$ [10] and $r(E)$  becomes energy independent equaling a trivial form: $-1$. Hence, no resonances are found.
This is the third instance (call it {\bf S0.3})  of smooth single-piece rising potential devoid of any resonances. In Fig. 3, we present the
cases of $h=0, g=1$ ({\bf S2.1}) and $h=1.2, g=1$. See small wiggles in $\tau(E)$ representing broad resonances. See Table 1 for the acclaimed closeness of $E_n$ and the the peak-energies, $\epsilon_n$, in $\tau(E)$. \\
\noindent
{\bf S3: Parabolic well joined to a constant potential} 
\begin{figure}[H]
\centering
\includegraphics{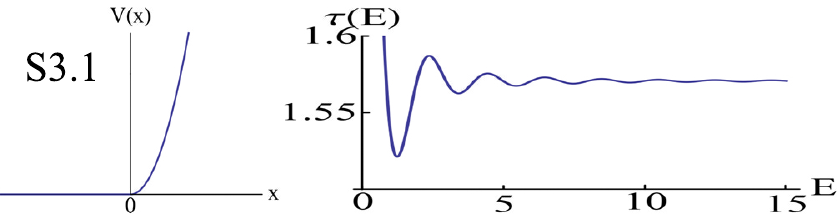}
\includegraphics{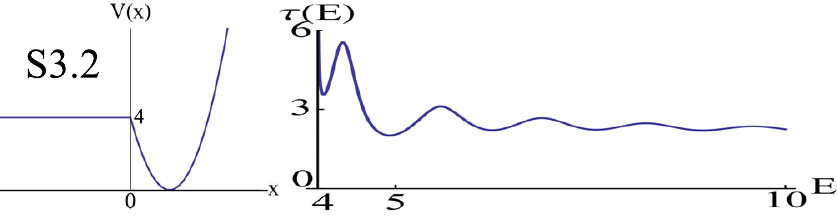}
\caption{The same as in FIG. 2 for system {\bf S3}; See Table 1.}
\end{figure}
\noindent
We can write this potential as
$V(x)=V_0=\frac{1}{2}m\omega^2a^2$, $x < 0$; $V(x)=\frac{1}{2}m\omega^2(x-a)^2$, $x \ge  0$. Using the parabolic cylindrical solution [2,9] of (1) for parabolic well, we can write $\psi(x)=C D_{\nu} (\beta (x-a))$ for $x>0$. When $x \rightarrow \infty$, $\psi(x) \sim  C^{\prime} (x-a)^{\nu} e^{-\beta^2(x-a)^2/4}\rightarrow 0 $. For $x<0$,  we seek $A e^{iqx}+ B e^{-iqx}$. Introducing $\nu=\frac{E}{\hbar\omega}-\frac{1}{2}$, we get
\begin{equation}
r(E)= \frac{-\beta D'_{\nu}(-\beta a ) + iqD_{\nu}(-\beta a )}{\beta D'_{\nu}(-\beta a) + iqD_{\nu}(-\beta a )}, ~ \beta = \sqrt{\frac{2m \omega}{\hbar}}.
\end{equation}
{\bf S4: Rising potential next to a step}\\
We consider $V(x)=-V'$, $x<0$; $V(x)=gx$, $x>0$ (see Fig.5).
Following the discussion of Airy function in ${\bf S2}$ above, the solution of (1) for this potential can be written as
$\psi(x)=A e^{ipx}+B e^{-ipx}$, $x \le 0$; $\psi(x)=C \mbox{Ai}(u(x))$, $x \ge 0$. In this case, we find
\begin{equation}
r(E)=\frac{ip\mbox{Ai} (\bar u) - g^{1/3} \mbox{Ai}'(\bar u)} {ip\mbox{Ai} (\bar u) + g^{1/3} \mbox{Ai}'(\bar u)},
\end{equation}
where $u(x), \bar u$ are the same as in {\bf S2} above.\\
See Fig. 5 for good peaks in time-delay and Table 1 for closeness  of ${E_n}$ and $\epsilon_n$. 
\begin{figure}[H]
\centering
\includegraphics{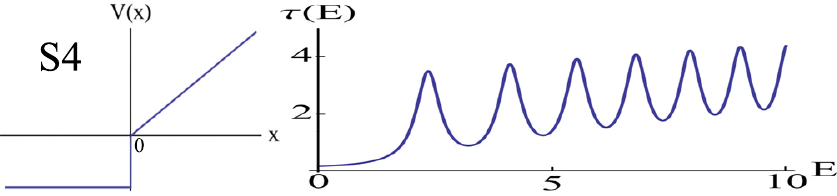}
\includegraphics{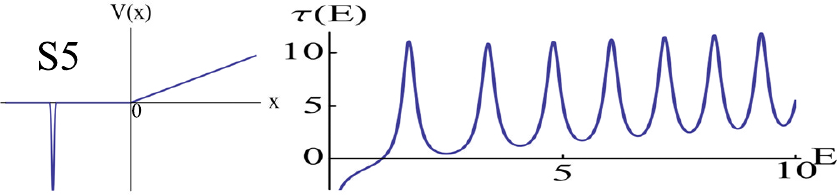}
\caption{The same as in FIG. 2 for {\bf S4} and {\bf S5}; see Table 1.}
\end{figure}
\noindent
{\bf S5: Rising potential next to a Dirac-delta well}
This system is represented by 
$V(x)=-V' \delta(x+a)$, $x<0$; $V(x)=g x$, $x \ge  0$.
Inserting this potential (see Fig. 5) in (1), the solution  can be written as
$\psi(x) = A e^{ikx}+B e^{-ikx}$, $x < -a$; $\psi(x)=C \sin kx + D\cos kx$, $-a < x \le 0$; $F \mbox{Ai}(u(x))$, $x > 0$. We find
\begin{eqnarray}
r(E) = e^{-2ika} \left( \frac{(-V' + ik)(-w\mbox{Ai}'\sin ka + k\mbox{Ai}\cos ka)- k(w\mbox{Ai}' \cos ka + k\mbox{Ai}\sin ka)}{(V' + ik)(-w\mbox{Ai}'\sin ka + k\mbox{Ai}\cos ka)
+ k(w\mbox{Ai}' \cos ka + k\mbox{Ai}\sin ka)} \right),~~
\end{eqnarray}
where $w=g^{1/3}$ and $\bar u$ is the argument of $\mbox{Ai}$ and $\mbox{Ai}'$. \\ \\
\noindent
{\bf S6: Rising potential next to a rectangular well}
\begin{figure}[H]
\centering
\includegraphics{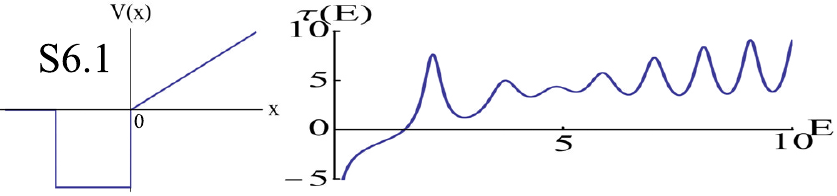}
\includegraphics{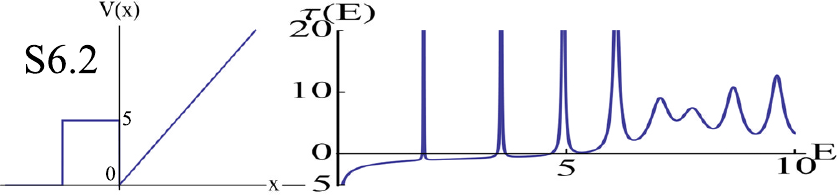}
\caption{ The same as in FIG. 2 for {\bf S6.1, S6.2.}
The system {\bf S6.2} possesses metastable states for $E<5$.
See Table 1.}
\end{figure}
\noindent
For this case, we have
$V(x)=0$, $x \le -a$; $-V'$, $-a<x<0$; $g x$, $x \ge  0$.
Fig 1(a) shows the schematic diagram of this system.
For this potential, the solutions of (1) can be written as $\psi(x)=A e^{ikx}+B e^{-ikx}$, $x \le -a$; $\psi(x)= C \sin px + D \cos px$, $-a <x <0$; $\psi(x)=F \mbox{Ai}(u(x))$, $x \ge 0$.
We find 
\begin{eqnarray}
r(E)=e^{-2ika} \left( \frac{ik(-w\mbox{Ai}'\sin pa+ p\mbox{Ai}\cos pa)-p(w\mbox{Ai}'\sin pa + p\mbox{Ai}\cos pa)}
{ik(-w\mbox{Ai}'\sin pa + p\mbox{Ai}\cos pa) 
+p(w\mbox{Ai}'\sin pa + p\mbox{Ai}\cos pa)} \right)
\end{eqnarray}
See Fig. 6 for $V(x)$ and $\tau(E)$  
\noindent
\begin{figure}[H]
\centering
\includegraphics[width=15 cm,height=5.5 cm]{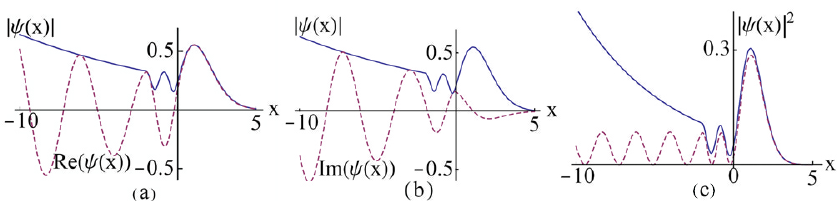}
\caption{Depiction of spatial catastrophe in eigen function at the first resonant energy of {\bf S6.1}-- (a) and (b) are for  $E = {\cal E}_1 = 2.1263 - 0.2428 i$. The solid line in (c) is for $E = {\cal E}_1$ and the dotted line for $E = \Re({\cal E}_1) = 2.1263$. $ |\psi(x)|^2$ for any energy, $E > 0$ (above the well, if any), come out to be  similar for any of these rising potentials ({\bf S0-S7}).}
\end{figure}
\noindent
{\bf S7: Rising potential next to a Gaussian well}

\begin{table*} [t]
\caption{First five resonances in various systems. ${\cal E}_n = E_n - i\Gamma_n/2 ~(\Gamma_n > 0)$ are the poles of $r(E)$ and $\epsilon_n$ are the peak positions in time-delay, $\tau(E)$. We take $2m = 1 =\hbar^2$. Notice the general closeness of $E_n$ and $\epsilon_n $, excepting the case of {\bf S3.1}.}
\begin{ruledtabular}
\begin{tabular}{|c||c||c||c||c||c||c||c||c|}
		\hline
		System & Fig. & Eq. & Parameters & ${\cal E}_0 (\epsilon_0)$ & ${\cal E}_1 (\epsilon_1)$ & ${\cal E}_2 (\epsilon_2)$ & ${\cal E}_3 (\epsilon_3)$ & ${\cal E}_4 (\epsilon_4)$\\
		\hline
		\hline
		S1.1 & 2 & (10) & $V_0=5,c=0.5$ & $8.88 - 1.50 i$ & $13.14 - 1.87 i$ & $17.30 - 2.17 i$ & $21.51 - 2.45 i$ & $25.80 - 2.70 i$ \\
	& & &  $d=5$& (8.89) & (13.21) & (17.34) & (21.65) & (26.05)\\
	\hline
		S1.2 & 2 & (10) &  $c = 0, d=5$ & $9.42 - 1.23 i$ & $13.77 - 1.49 i$ & $18.01 - 1.69 i$ & $22.28 - 1.89 i$ & $26.62 - 2.07 i$ \\
	 & & & $V_0=5$ & $(9.36)$ & $(13.46)$ & $(18.04)$ & $(22.14)$ & $(26.43)$\\
	\hline	
	S2.1 & 3 & (11) & $h = 0, g = 1$ & $2.79 - 1.09 i$ & $4.45 - 1.03 i$ & $5.85 - 0.98 i$ & $7.08 - 0.94 i$ & $8.22 - 0.91 i$ \\
	&  & & & $(2.89)$ & $(4.62)$ & $(5.85)$ & $(7.12)$ & $(8.26)$\\
	\hline	
	S2.2 & 3 & (11) & $h = 1.2, g = 1$ & $1.88 - 1.69 i$ & $3.73 - 1.47 i$ & $5.21 - 1.35 i$ & $6.50 - 1.27 i$ & $7.68 - 1.22 i$ \\
	&  & & & $(1.92)$ & $(3.98)$ & $(5.24)$ & $(6.53)$ & $(7.76)$\\
	\hline		
	S3.1 & 4 & (12) &  $a = 0, ~\hbar \omega = 1$ & $2.08 - 1.75 i$ & $4.21 - 2.09 i$ & $6.28 - 2.32 i$ & $8.32 - 2.48 i$ & $10.34 - 2.61 i$ \\
	 & & & & $(2.38)$ & $(4.43)$ & $(6.51)$ & $(8.54)$ & $(10.33)$\\
	\hline	
	S3.2 & 4 & (12) & $a = 4, ~\hbar \omega = 1$ & $4.32 - 0.22 i$ & $5.56 - 0.48 i$ & $6.86 - 0.65 i$ & $8.20 - 0.78 i$ & $9.58 - 0.89 i$ \\
	&  & & & $(4.31)$ & $(5.56)$ & $(6.86)$ & $(8.21)$ & $(9.63)$\\
	\hline	
		S4 & 5 & (13) &  $V' = 10, g = 1$ & $2.34 - 0.31 i$ & $4.09 - 0.30 i$ & $5.52 - 0.29 i$ & $6.79 - 0.29 i$ & $7.95- 0.28 i$ \\
	&  & & & $(2.31)$ & $(3.96)$ & $(5.53)$ & $(6.79)$ & $(7.90)$\\
	\hline
	S5 & 5 &(14) & $V' = 2, a = 1$ & $1.69 - 0.17 i$ & $3.39 - 0.18 i$ & $4.79 - 0.18 i$ & $6.04 - 0.18 i$ & $7.19- 0.18 i$ \\
	 & & & $g= 1$& $(1.69)$ & $(3.42)$ & $(4.82)$ & $(6.06)$ & $(7.20)$\\
	\hline	
	S6.1 & 6 & (15) &  $V' = 5, a = 2$ & $2.13 - 0.24 i$ & $3.69 - 0.43 i$ & $4.81 - 0.57 i$ & $5.85 - 0.43 i$ & $6.98 - 0.32 i$ \\
	 & & & $g= 1$& $(2.15)$ & $(3.73)$ & $(4.78)$ & $(5.84)$ & $(7.06)$\\
	\hline	
	S6.2 & 6 & (15) &  $V' = -5, a = 2$ & $1.86 - 0.5\times 10^{-3}i$ & $3.56 -0.3 \times 10^{-3} i$ & $4.93 - 0.01 i$ & $6.08-0.07 i$ & $7.04 - 0.25 i$ \\
	 & & & $g= 1$& $(1.85)$ & $(3.56)$ & $(4.95)$ & $(6.10)$ & $(7.03)$\\
	\hline	
\end{tabular}
\end{ruledtabular}
\end{table*}
\noindent
We write this potential as $V(x)=-V_0 e^{-(x+a)^2/b^2}$, $x<0$; $V(x)= gx$, $x\ge0$. We find its reflection amplitude and time- delay by numerically integrating Eq. (1). To testify this method, we have reproduced the results of {\bf S6} (see Eq. (15)). 
The numerically solved example of a Gaussian potential well juxtaposed
to a linear rising potential (Fig.7 ) giving rise to multiple peaks in time-delay speaks
of the good universality of resonances in two-piece rising potentials.
\begin{figure}[H]
\centering
\includegraphics{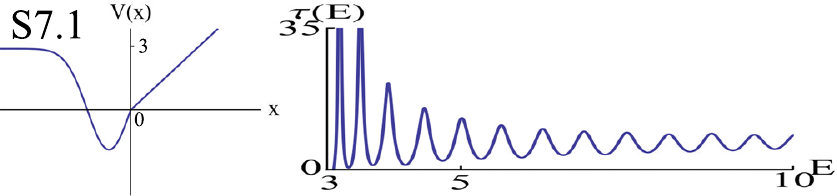}
\includegraphics{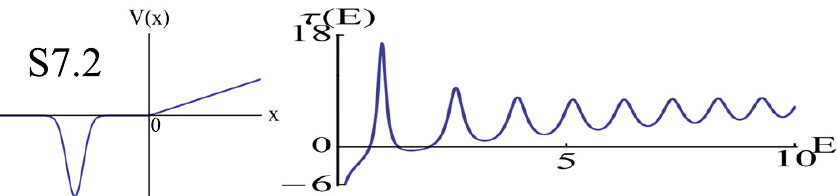}
\includegraphics{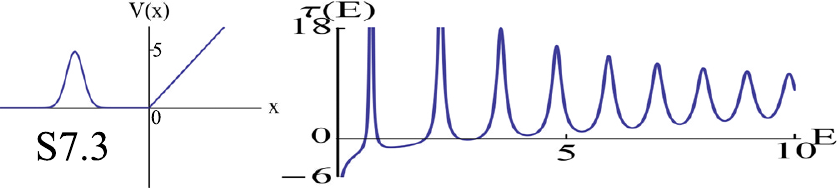}
\caption{The same as in FIG. 2 for system {\bf S7}. For ${\bf S7.1}$: $V'=5$, $a=1$, $b=\sqrt{2}$, $g=1$. Below $E=3$, there are both positive and negative energy bound states (not in the graph).  For {\bf S7.2}: $V'=10$, $a=2$, $b=.31$, $g=1.$ For {\bf S7.3}: $V'=-5$, $a=2$, $b=.31$, $g=1$.
For $E<5$ ($E>5)$ the peaks are for metastable (resonance) states.}
\end{figure}
Most remarkably, the other piece of the potential (for $x<0$) could  just be zero, a step or a well or a barrier.
Moreover, the well itself may nor may not have bound states or resonances. All ({\bf S1} to {\bf S7}) show resonances which may be both broad (see {\bf S1-S3}) and sharp (see {\bf S4-S7}).
Though not presented here, we find that a rising potential attached to more than two piece potentials also yield resonances. We have shown linear potential in systems ({\bf S4-S7}) as rising parts, this being easily achievable as a constant electric field for $x>0$. Parabolic and exponential  profiles also give the same effects(resonances as peaks in time-delay) qualitatively. The values of first five complex energy poles (${\cal E}_n$) of $r(E)$ and the peak positions ($\epsilon_n)$ of time-delay, compiled in Table 1 for various systems, show a good agreement except in the curious case of {\bf S3.1}.
For the typical catastrophic behavior of $\psi(x)$ for these systems at $E={\cal E}_0, E_0$, and any other positive energy not within the well (if any), see Fig. 7.

The systems {\bf S6.2}  and {\bf S7.3} have been presented here for both metastable states below the barrier height  and resonances above it; see Table 1 for details. In these cases, if we increase the value of $g$, the rising part tends to become a hard (rigid) wall; Gamow's complex energy states become fewer and broader. Hence, we conclude that the strength of rising potential is crucial for controlling the number and quality of resonances in the new systems. Orthodox one-dimensional potentials with three real turning points   may have paucity of resonances but they are found easily in rising potentials like $V_{\lambda}(x)$ [1] and the two-piece systems presented here. 

We hope that the absence of resonances in rising one-piece potentials (Fig. 1) has been well noted; a proof for/against the same is welcome. We have attributed the occurrence of resonances to rising potentials and to their two-piece nature. In this regard, the odd parabolic potential, $x|x|$ [4], is a curious example having only one peak in time delay and no catastrophe in the  eigenfunction at $E={\cal E}_0$. The cubic potential $-x^3$ is known [4] to have non-real discrete spectrum. However, it would be interesting to carry out the time delay analysis for this and other odd potentials like $x^{2n+1}$ and $x^{2n}|x| (n=1,2,3,..$) to see if  they compete with simple potentials presented here for peaks in time delay. It needs to be investigated if there exists a continuous two-piece rising potential (saturating towards $x\sim -\infty$)  devoid of even wiggles in time-delay corresponding to ${\cal E}_n$ ($\Gamma_n > 0 $) in  all of its parametric regimes. Search for (the hitherto elusive) necessary condition for occurrence of resonances in a potential has now become more intriguing.

The intimate connection of quantum mechanics with (electromagnetic and matter) wave-propagation [15] ensures that the new avenue of resonances will be useful there as well, may be even more. The experimental verification and harnessing of the new avenue of resonances is the most pressing question arising here. In this regard, we hope that the system {\bf S2.1} presented above serves as a simple, feasible model which can be designed in laboratories easily; it is like a charged particle being injected in a region of constant electric field.

\end{document}